\documentclass[a4paper,twocolumn,showpacs]{revtex4}

\usepackage{amsmath}
\usepackage{graphicx}
\usepackage[english]{babel} 
\usepackage[latin1]{inputenc} 
\usepackage[T1]{fontenc} 

\begin{document}
\title{Generalized Manna sandpile model with height restrictions}
\author{Wellington Gomes Dantas}
\email{wgd@if.uff.br}
\author{J\"urgen F. Stilck}
\email{jstilck@if.uff.br}
\affiliation{Instituto de F\'{\i}sica\\
Universidade Federal Fluminense\\
Av. Litor\^anea s/n\\
24210-340 - Niter\'oi, RJ\\
Brazil}
\date{\today}

\begin{abstract}
Sandpile models with conserved number of particles (also called fixed energy
sandpiles) may undergo phase transitions between active and absorbing states.
We generalize the Manna sandpile model with fixed number of particles,
introducing a parameter $-1 \leq \lambda \leq 1$ related to the toppling of
particles from 
active sites to its first neighbors. In particular, we discuss a model with
height restrictions, allowing for at most two particles on a site. Sites
with double occupancy are active, and their particles may be transfered to
first neighbor sites, if the height restriction do allow the change. For
$\lambda=0$ each one of the two 
particles is independently assigned to one of the two first neighbors and the
original stochastic sandpile model is recovered. For $\lambda=1$ exactly one
particle will be placed on each first neighbor and thus a deterministic (BTW)
sandpile model is obtained. When $\lambda=-1$ two particles are moved to one
of the first neighbors, and this implies that the density of active sites is
conserved in the evolution of the system, and no phase transition is
observed. Through simulations of the stationary state, we estimate the
critical density 
of particles and the critical exponents as functions of $\lambda$.
\end{abstract}

\pacs{05.70.Ln, 02.50.Ga, 64.60.Cn,64.60.Kw}

\maketitle

\section{Introduction}
\label{intro}
Problems related to phase transitions between active and absorbing states
have attracted much interest in recent years \cite{h00}.
Although these transitions occur away from thermodynamic equilibrium, since the
presence of absorbing states prevents detailed balance to be satisfied in the
dynamical evolution, the theoretical framework developed for
equilibrium phase transition may in fact be applied to these systems, and
concepts such as scaling and universality are relevant for non-equilibrium
phase transitions as well. It is then of interest to identify the
universality classes in these systems. Many of the stochastic models with
absorbing states which show a phase transition belong to the directed
percolation (DP) universality class \cite{md99}, including the much studied
contact process (CP) \cite{h74}. The so called DP conjecture \cite{j81},
confirmed so far through many examples, states that models with a scalar order
parameter exhibiting phase transitions between an active state to a single
absorbing state and without additional conservation laws should belong to this
universality class. In models for 
sandpiles the number of particles is conserved and an infinite number of
absorbing 
states is present, and thus they are potential candidates to belong to a non-DP
universality class. Careful simulations of the unrestricted Manna
sandpile model \cite{d01} and the model with height restrictions
\cite{dtm02} lead to exponents at variance with the DP value.

The Manna sandpile model was originally proposed as a stochastic model for
SOC, with slow addition of sand and avalanches which lead to abrupt loss of
sand \cite{m90}. The scaling behavior in the SOC regime was later recognized
to be associated to an absorbing state phase transition in the corresponding
model without addition or loss of particles (grains of sand), the so called
fixed energy sandpiles (FES) \cite{vz97}. In this model, a $d$-dimensional
lattice of size $L^d$ is occupied by $N$ particles. A configuration of the
lattice is specified fixing the number of particles $z_i$ at each site $i$,
where $z_i$  may be any non-negative integer number. Sites with $z_i \geq 2$
are active, and an active site may lose two particles to its first neighbors,
with a unitary toppling rate. The two particles move to randomly and
independently chosen sites among the first neighbors of site $i$. Any state in
which no site has two or more particles is an absorbing state, and in the
thermodynamic limit the number of absorbing states is infinite, as long as the
density of particle $\zeta=N/L^d$ is smaller than unity. Numerical simulations
\cite{d01} in fact show that this model in one dimension undergoes a
continuous phase transition at a critical density $\zeta_c=0.9488$ with
critical exponents which are different from the DP values.

In this paper we study a variation of the original model where the occupancy
numbers are restricted to $z_i \leq 2$. Each move of a particle is accepted
only 
if this constraint is not violated at the destination site. This additional
restriction in the allowed configurations leads to simplifications in
mean-field calculations and in simulations. Numerical simulations for the
one-dimensional 
restricted model lead to a slightly lower critical density $\zeta_c=0.92965$,
while the exponents are compatible with the ones found for the unrestricted
model \cite{dtm02}. We generalize this restricted model by introducing a
parameter whose value is related to the choice of the destination sites of the
two toppling particles. The restricted Manna model corresponds to $\lambda=0$
and a model where exactly one particle is sent to each first neighbor of the
active site (in one dimension) is recovered for $\lambda=1$. We call this
latter case the BTW model, since it corresponds to the original Bak, Tang and
Wiesenfeld model \cite{btw87} with conserved number of particles. In
particular, we are interested in investigating the phase transition in the
model as the parameter $\lambda$ is changed.

In the section \ref{model} we define the model in more details. The numerical
simulations which were performed are described in section \ref{sim}, as well
as the results they furnished. Conclusions may be found in section \ref{conc}.

\section{Definition of the model}
\label{model}
We consider $N$ particles located on the $L$ sites of a one-dimensional
lattice in such a way that each site $i$ is occupied by $z_i=0,1$ or $2$
particles. Sites with two particles are \emph{active} and the control
parameter of the model is the density of particles $\zeta=N/L$. The update of
the configuration is started by randomly choosing one of the $N_A$ active site
of the lattice. The following updates of the configuration may then occur:
\begin{enumerate}
\item With a probability $(1+\lambda)/2$ each first neighbor receives exactly
one of the particles from the active site, as may be seen in figure
\ref{trans1}.

\begin{figure}[h!]
\vspace{0.3cm}
\begin{center}
\includegraphics[height=1.5cm]{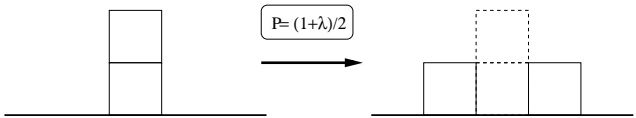}
\caption{The two particles of the active sites are moved to different first
neighbor sites.}
\label{trans1}
\end{center}
\end{figure}

If one of the first neighbors is active, the particle which was chosen to be
sent to this site remains on the original site. If both first neighbor sites
are 
active, no particle movement occurs. This assures that the restriction on the
occupancy numbers is always satisfied.

\item With a probability $(1-\lambda)/2$, one of the two first neighbors is
chosen with probability 1/2 and both particles are moved to this site. If the
destination site is already occupied by one particle, only one of the
particles originating from the active site is moved. If the destination site
is active, no movement is done. An example of such a transition is shown in
figure \ref{trans2}.

\begin{figure}[h!]
\begin{center}
\includegraphics[height=1.5cm]{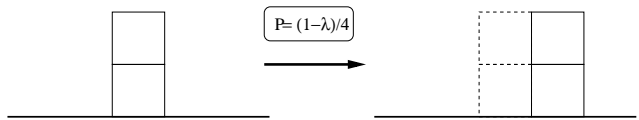}
\caption{Both particles from the active site are moved to the same first
neighbor site.}
\label{trans2}
\end{center}
\end{figure}
\end{enumerate}

One iteration as described above corresponds to a time increment  of
$1/N_A$. After each iteration, the list of active sites is updated. In the
restricted model the density of particles obeys $\zeta \leq 2$, and when
$\zeta=2$ an additional absorbing state is reached. In the unrestricted model 
there is no such upper limit for
$\zeta$ and no spurious absorbing state with unitary density of active sites
exists. However, since the phase transition in the models occurs at particle
densities below unity, this details do not bother us here. The fact that the
restriction on the occupancy numbers implies a reduction of only about 2\% in
the critical density of the model reveals that for the unrestricted model in
the supercritical region close to the phase transition the fraction of sites
with occupancy larger than 2 should be very small.

The parameter $\lambda$ is restricted to the interval $[-1,1]$. In the upper
limit of this interval, the toppling of particles is deterministic and the
model corresponds to a BTW sandpile with conserved number of particles. In the
lower limit the number of active sites is a conserved quantity in the time
evolution of the model, with active sites diffusing on the lattice. The
conservation of the number of active sites in this limit is a consequence of
the occupancy numbers restriction and does not happen in the unrestricted
model. 

\section{Simulations and results}
\label{sim}
We realized simulations to find the critical properties of the one-dimensional
model with height restrictions. The initial condition, for a given value of
$\lambda$ and $\zeta=N/L$, is a uniform and uncorrelated distribution of the
$N$ particles on the $L$ sites of the lattice, respecting the restrictions. We
studied lattice sizes $L$ between 100 and 2000, performing $N_r$ repetitions
with times up to $t_{max}$ for a certain range of densities and each value of
$\lambda$. In our simulations $t_{max}$ was in the interval $[4 \times 10^4,
2 \times 10^7]$ and $N_r=2000$. Since in finite systems with absorbing states
the only stationary states are the absorbing states themselves, to study the
transition we are interested in the quasi-stationary states of the
model. Usually, the simulational determination of this state is hindered by
the presence of the absorbing states, since considerable fluctuations will be
found in the quantities estimated in the simulations, particularly for
particle densities close to the critical value. To avoid these drawbacks and
increase the precision of the estimates of the stationary state we used the
prescription for simulation of quasi-stationary states proposed recently by
de Oliveira and Dickman \cite{od05}.

\begin{figure}[h!]
\vspace{0.8cm}
\begin{center}
\includegraphics[height=5.5cm]{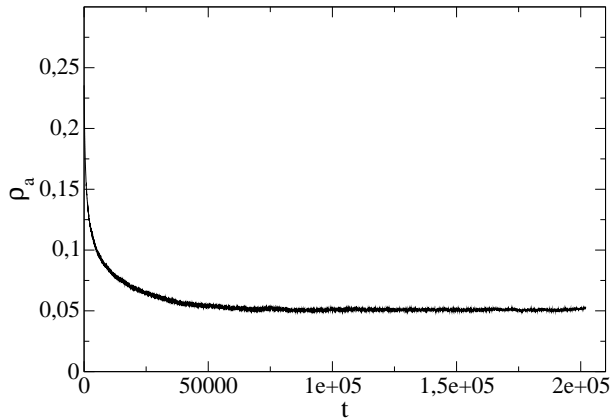}
\caption{Example of a simulation of the time evolution of the density of
active sites. In this case $L=100$, $\lambda=0$, and $\zeta=0.92$.}
\label{te}
\end{center}
\end{figure}

In the simulations, the time evolution of the density of active sites $\rho_a$
is obtained, one example being shown in figure \ref{te}. A stationary value of
this density may then be estimated for given values of $\zeta$ and $\lambda$
calculating the mean value of the last 2000 points in such simulations.

\begin{figure}[h!]
\vspace{0.8cm}
\begin{center}
\includegraphics[height=5.5cm]{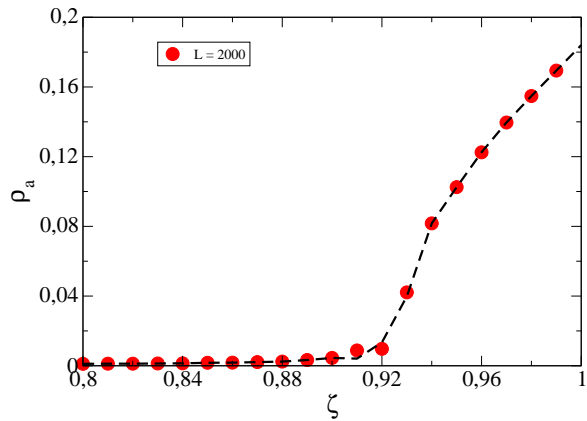}
\caption{Estimated values of the order parameter $\rho_a$ as a function of the
particle density $\zeta$, obtained for $L=2000$ and $\lambda=0$.}
\label{op}
\end{center}
\end{figure}

For each value of the lattice size $L$, realizing simulations for a range of
particle densities $\zeta$, we obtain curves for the order parameter such as
the one depicted in figure \ref{op}. It should be mentioned that for each
lattice size the particle density may assume only a discrete set of values,
but it seems reasonable to interpolate between those values, assuming that
$\rho_a$ varies continuously with $\zeta$. 

The estimated values of the order parameter as a function of the particle
density $\zeta$ may then be analyzed considering the following scaling
relations:
\begin{eqnarray}
\rho_a(\zeta_c,L) &\sim& L^{-\beta/\nu_{\perp}},\\
\rho_a(\zeta) &\sim& (\zeta-\zeta_c)^{\beta}.
\end{eqnarray}
Here $\rho_a(\zeta)$ denotes the thermodynamic limit $L \to \infty$ of
$\rho_a(\zeta,L)$. These scaling relations provide estimates for the critical
exponents $\beta$ and $\nu_\perp$, as well as for the critical particle
density $\zeta_c$. To estimate the critical particle density and the order
parameter exponent $\beta$ we proceed as follows: for a given value of $L$ we
choose a critical value $\zeta_c^{(L)}$ which maximizes the correlation in a
linear approximation for the function $\ln \rho_a = b+
\beta^{(L)}\ln[\zeta-\zeta_c^{(L)}]$, with $\zeta>\zeta_c^{(L)}$. 
Once obtaining
estimates for $\beta^{(L)}$ and $\zeta_c^{(L)}$ for different values of $L$,
they may be extrapolated to the limit $L \to \infty$. The ratio
$\beta/\nu_\perp$ may be estimated directly through the relation $\ln
\rho_a=c+\beta/\nu_\perp \; \ln L$, using values for $\rho_a$ at the estimated
critical particle density.

\begin{figure}[h!]
\vspace{0.5cm}
\begin{center}
\includegraphics[height=5.2cm]{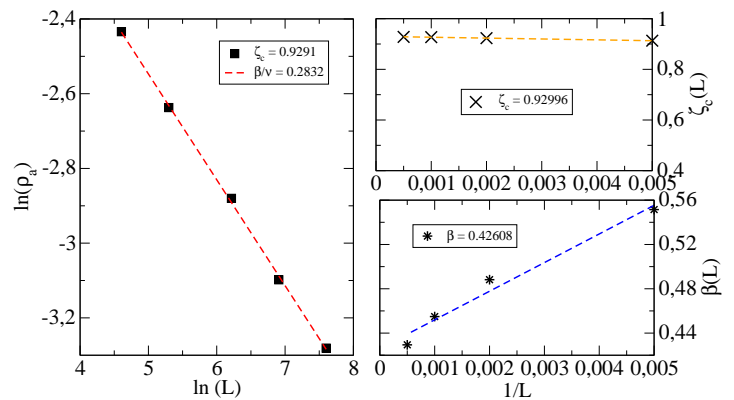}
\caption{Extrapolation without the scaling correction for $\lambda=0$. The
estimates are: $\zeta_c=0.92996$, $\beta/\nu_{\perp}=0.2832$ and
$\beta=0.42608$.} 
\label{cesc}
\end{center}
\end{figure}

\begin{figure}[h!]
\vspace{0.5cm}
\begin{center}
\includegraphics[height=5.cm]{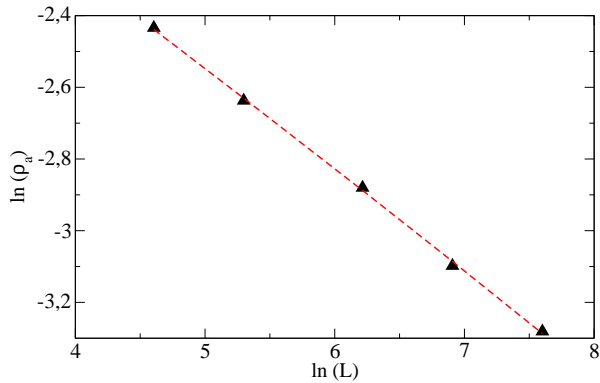}
\caption{Extrapolations with the scaling correction for  
$\lambda=0$. Corrected values for the estimates:
$\zeta_c=0.92996$, $\beta/\nu_{\perp}=0.247$ and $\beta=0.41275$.} 
\label{lamb0}
\end{center}
\end{figure}

Using the procedure described above, we obtained estimates for the case
$\lambda=0$, as may be seen in figure \ref{cesc}. We then realized that the
estimates for $\beta$ may be improved excluding the smaller sizes in the
extrapolation to the thermodynamic limit. A better result for the ratio
$\beta/\nu_{\perp}$ is found if we apply the scaling correction proposed in
\cite{od05} for the 
simulation of the quasi-stationary state of the contact process. It seems that
the algorithm proposed in this reference for simulations of the
quasi-stationary state and used in our work implies a scaling correction which
is characteristic of the model. With these corrections, our estimates are very
close to the ones in reference \cite{dtm02}. The extrapolations with the
scaling correction may be seen in figure \ref{lamb0}.

\begin{figure}[h!]
\vspace{0.5cm}
\begin{center}
\includegraphics[height=5.5cm]{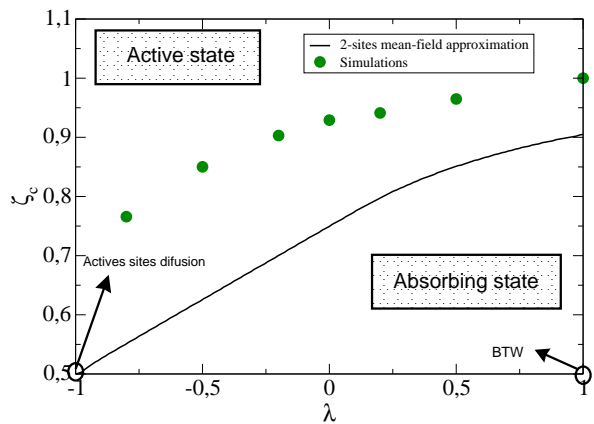}
\caption{Phase diagram of the Manna model with height restriction parametrized
by $\lambda$.}
\label{cl}
\end{center}
\end{figure}

Repeating this strategy for different values of $\lambda$, we finally may
obtain the phase diagram of the model, as well as study the values of the
estimates for the critical exponents as functions of this parameter. In figure
\ref{cl} the phase diagram is shown. Besides the points for the critical line
which emerge from the simulations, the result of a 2-site mean field
approximation for the model may also be seen in the figure. Details of these
calculations will be given elsewhere. The curve which results
from the mean-field calculations is always below the results of the
simulations, confirming that when the correlations are ignored the size of the
active region in the phase diagram is overestimated. As $\lambda$ approaches
the limiting value -1, the results of the simulations get closer to the ones
in the mean-field approximation, which leads to $\zeta_c=1/2$ in this limit,
even in a one-site approximation. This seems to be reasonable, since in this
limit the evolution of the model is dominated by the diffusion of active
sites, and this leads to a mean-field like behavior. This argument applies
only in the limit of $\lambda$ slightly above -1, since as already noticed
above when $\lambda=-1$, $\rho_a(t)$ is constant and no phase transition is
observed. 

\begin{figure}[h!]
\vspace{0.9cm}
\begin{center}
\includegraphics[height=5.cm]{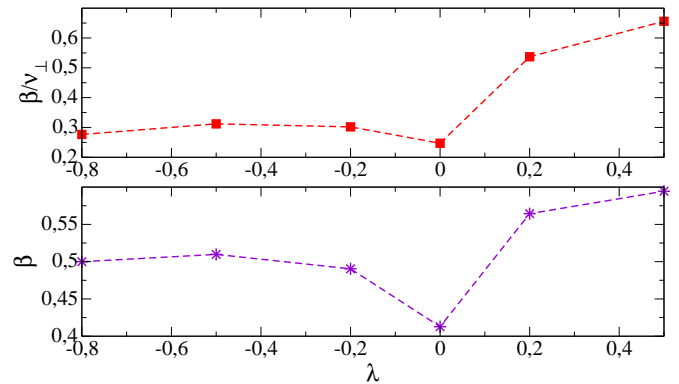}
\caption{Estimates for the critical exponents $\beta$ e $\beta/\nu_{\perp}$ as
functions of the parameter $\lambda$.}
\label{cexps}
\end{center}
\end{figure}

\begin{figure}[h!]
\vspace{0.8cm}
\begin{center}
\includegraphics[height=5.cm]{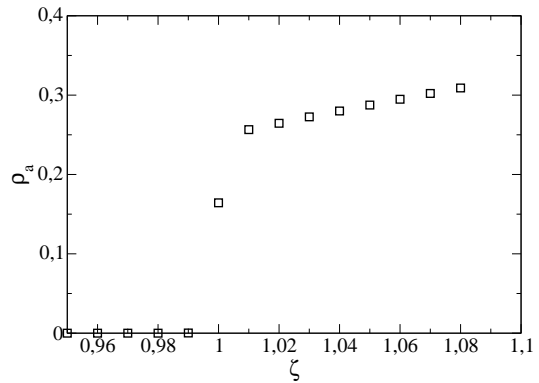}
\caption{Estimates for the order parameter for $L=500$, with $\lambda=1$,
indicating a first order transition.}
\label{fo}
\end{center}
\end{figure}

In figure \ref{cexps} we notice that the estimates for the critical exponents
show a rather strong dependence of the parameter $\lambda$, which may be due
to a crossover between universality classes in the limits $\lambda \to -1$,
where mean-field behavior is expected,  and $\lambda \to 1$, the conservative
BTW model. In this latter limit, our simulations indicate a first-order
transition at $\zeta=1$, as may be seen in figure \ref{fo}

\section{Conclusion}
\label{conc}
We studied a generalization of the one-dimensional Manna model with height
restrictions and conservation of the number of particles, with the inclusion
of a parameter $\lambda$ which is related to the two toppling processes which
occur in the model. When $\lambda=-1$ no transition is found between an active
and an absorbing state, and a diffusive dynamics of active sites is found. In
the other extreme $\lambda=1$ the model corresponds to a conservative BTW
sandpile, and our results indicate a discontinuous transition. We believe that
the observed variations in the critical exponents may be due to crossover
effects in the two limiting cases of the model.

It is necessary to extend our simulations to values of $\lambda$ which are
closer to both limits, in order to find out if the exponents approach limiting
values. In particular, it would be interesting to find out if mean-field
exponents are found as $\lambda \to -1$. The exponent ration
$\nu_{\parallel}/\nu_{\perp}$ should also be estimated through additional
simulations, as well as the ratio $m=\langle\rho_a^2\rangle/\rho_a^2$, whose
value at $\zeta=\zeta_c$ is also universal. We believe the phase
transition in the model to be discontinuous only at $\lambda=-1$, but the data
we have collected so far does not allow us to discard the possibility that a
tricritical point exists for some value of $\lambda$ between 0 and 1.

We are presently trying to answer these questions, so that the critical
behavior of this model, which does not belong to the DP universality class,
may be better known.

\begin{acknowledgments}
We thank Prof. Ronald Dickman for many helpful discussions. This
research was partially supported by the Brazilian agencies CAPES, FAPERJ
and CNPq, whose assistance is gratefully acknowledged. 
\end{acknowledgments}

\end{document}